\documentclass[a4paper,11pt]{article}
\usepackage{graphicx}
\usepackage{amssymb}
\usepackage{amsmath}
\usepackage{mathtools}
\usepackage{amsfonts}
\usepackage{appendix}
\usepackage{color}
\usepackage{multirow}
\usepackage{pdflscape}
\usepackage{caption}
\usepackage{subcaption}
\usepackage{xcolor}
\usepackage{setspace}
\usepackage{booktabs}
\usepackage{adjustbox}
\def\diag{\mathop{\rm diag}\nolimits}

\def\var{\mathop{\rm Var}\nolimits}

\newcommand{\bbeta}{\boldsymbol{\beta}}
\newcommand{\bbhat}{\boldsymbol{\widehat{\beta}}}

\newcommand{\bdelta}{\boldsymbol{\delta}}

\newcommand{\0}{\boldsymbol{0}}

\newcommand{\bphi}{\boldsymbol{\phi}}

\newcommand{\bb}{\boldsymbol{b}}

\newcommand{\bC}{\boldsymbol{C}}

\newcommand{\bD}{\boldsymbol{D}}
\newcommand{\be}{\boldsymbol{e}}

\newcommand{\bF}{\boldsymbol{F}}

\newcommand{\bI}{\boldsymbol{I}}

\newcommand{\bR}{\boldsymbol{R}}

\newcommand{\bu}{\boldsymbol{u}}

\newcommand{\bv}{\boldsymbol{v}}

\newcommand{\bW}{\boldsymbol{W}}
\newcommand{\bx}{\boldsymbol{x}}
\newcommand{\bX}{\boldsymbol{X}}

\newcommand{\bY}{\boldsymbol{Y}}

\newcommand{\calK}{\mathcal{K}}
\newcommand{\calL}{\mathcal{L}}

\newcommand{\calN}{\mathcal{N}}

\def\diag{\mathop{\rm diag}\nolimits}

\def\var{\mathop{\rm Var}\nolimits}

\newtheorem{theorem}{Theorem}

\newtheorem{proposition}[theorem]{Proposition}

\begin{document}

\title{\bf LLASSO: A linear unified LASSO for multicollinear situations}

\bigskip

\author{{ M. Arashi$^{1}$\footnote{Corresponding Author, Email:m\_arashi\_stat@yahoo.com}, Y. Asar$^{2}$ and B. Y\"{u}zba\c{s}{\i}$^{3}$}
\vspace{.5cm} \\\it$^{1}$ Department of Statistics, School of
Mathematical Sciences\\\vspace{.5cm} \it Shahrood University of Technology, Shahrood, Iran \\
\it$^{2}$ Department of Mathematics-Computer Sciences,\\\vspace{.5cm} 
\it Necmettin Erbakan University, Konya, Turkey\\
\it$^{3}$ Department of Econometrics, Inonu University, Malatya, Turkey
 }

\date{}
\maketitle

\begin{quotation}
\noindent {\it Abstract:} We propose a rescaled LASSO, by premultipying the LASSO with a matrix term, namely linear unified LASSO (LLASSO) for multicollinear situations. Our numerical study has shown that the LLASSO is comparable with other sparse modeling techniques and often outperforms the LASSO and elastic net. Our findings open new visions about using the LASSO still for sparse modeling and variable selection. We conclude our study by pointing that the LLASSO can be solved by the same efficient algorithm for solving the LASSO and suggest to follow the same construction technique for other penalized estimators.

\par

\vspace{9pt} \noindent {\it Key words:} Biasing parameter; $l_1$-penalty; LASSO; Liu Estimation; Multicollinearity; Variable selection.

%\par

%\vspace{9pt} \noindent {\it AMS Classification:} Primary: 62F15,
%Secondary: 62H05\par

\end{quotation}\par

\section{Introduction}
Let $\{(\bx_i,Y_i),i=1,\ldots,n\}$ be a random sample from the linear regression model
\begin{equation}\label{eq11}
Y_i=\bx_i^{\top}\bbeta+\epsilon_i
\end{equation}
where $Y_i\in\mathbb{R}$ is the response, $\bx_i=(x_{i1},\ldots,x_{ip})^{\top}\in\mathbb{R}^p$ is the covariate vector and $\epsilon_i$ is the random error with $\mathbb{E}(\epsilon_i)=0$, $\var(\epsilon_i)=\sigma^2\in\mathbb{R}^+$.

The ordinary least squares (OLS) estimator has the form $\hat\bbeta_n=\bC_n^{-1}\bX^{\top}\bY$, $\bC_n=\bX^{\top}\bX$.
%\begin{equation}\label{OLS}
%\hat\bbeta_n=\bC_n^{-1}\bX^{\top}\bY,\quad\bC_n=\bX^{\top}\bX
%\end{equation}
For the high-dimensional case $(p>n)$, the OLS estimator is not valid, and in this case one may use a regularization method to find a few non-zero elements of $\bbeta$, as a remedial approach. Under the $\textit{l}_1$-penalty, Tibshirani (1996) proposed the least absolute penalty and selection operator (LASSO) given by
\begin{equation}\label{LASSO}
\hat\bbeta_n^{L}=\arg\min_{\bbeta}\left\{\frac{1}{n}\|\bY-\bX\bbeta\|^2_2+\lambda\|\bbeta\|_1\right\},
\end{equation}
where $\bY=(Y_1,\ldots,Y_n)^{\top}$, $\lambda>0$ is the threshold, and $\|\bv\|_q=(\sum_{j=1}^d |v_j|^q)^{1/q}$ for $\bv=(v_1,\ldots,v_d)^{\top}$, with $q>0$.

The LASSO has tractable theoretical and computational properties. However, when the predictors $\bx_i$ are highly correlated, the LASSO may contain too many zeros. This is not undesirable, but it may have some effects on prediction. Refer to Zou and Hastie (2005) for limitations of LASSO. As a remedy, one may use projection pursuit with the LASSO or apply the well-known ridge regression (RR) estimator of Hoerl and Kennard (1970). Unlike LASSO, the RR estimator does not ``kill" coefficients and hence it cannot be used as an efficient estimator in sparse models.
Zou and Hastie (2005) introduced the Elastic Net (E-net) approach which can deal with the strongly correlated variables effectively. Like LASSO, the E-net has also some promising properties. The E-net is given by
\begin{equation}\label{E-net}
\hat\bbeta_n^{En}=\arg\min_{\bbeta}\left\{\frac{1}{n}\|\bY-\bX\bbeta\|^2_2+\lambda_2\|\bbeta\|_2^2+\lambda_1\|\bbeta\|_1\right\},
\end{equation}
where $\lambda_1$ and $\lambda_2$ are non-negative tuning parameters.

Indeed the E-net is an improved LASSO, which the penalty of ridge approach is taken into account in the optimization problem. Zou and Hastie (2005) formulated the na\"{i}ve E-net in such a way the solution to the optimization problem connected with that of the LASSO. In the same line, we have different concern which is motivated in below.

\subsection{Motivation}
Under a multicollinear situation, apart from the sparsity, the OLS estimator $\hat\bbeta_n$ is far away from the true value $\bbeta$. Hence, it is of major importance to find a closer estimator. Based on the Tikhonov's (1963) regularization approach, Hoerl and Kennard (1970) proposed to minimize the sum of squares error (SSE) subject to $\|\bbeta\|_2^2=k$, to obtain the RR estimator. The RR estimator is a non-linear function with respect to the tuning (biasing, here) parameter, in nature. Another approach to combat multicollinearity is to minimize the SSE subject to $\|d\hat\bbeta_n-\bbeta\|_2^2=k$, $0<d<1$, due to Mayer and Willke (1973). The idea is $d\hat\bbeta_n$ is closer to the true value $\bbeta$ for the case $0<d<1$, than $\hat\bbeta_n$. The resulting estimator is linear unified (Liu) estimator $\bF_n(d)\hat\bbeta_n$
%\begin{equation*}
%\hat\bbeta_n(d)=(\bC_n+\bI_{p})^{-1}(\bX^{\top}\bY+d\hat\bbeta_n)=\bF_n(d)\hat\bbeta_n
%\end{equation*}
where $0<d<1$ is the biasing parameter and $\bF_n(d)=(\bC_n+\bI_p)^{-1}(\bC_n+d\bI_p)$ is the biasing factor. Apparently, the Liu estimator is linear with respect to the biasing parameter $d$. Note that, in contrast with this estimator, the RR estimator has the form $\bR_n(k)\hat\bbeta_n$, $\bR_n(k)=(\bI_p+k\bC_n^{-1})^{-1}$, with $k>0$.

The key idea in our approach is to make use of this difference between $\bR_n(k)\hat\bbeta_n$ and $\bF_n(d)\hat\bbeta_n$ in obtaining a better estimator. Hence, we propose to replace the penalty term $\lambda_2\|\bbeta\|_2^2$ in the E-net by $\lambda_2\|d\hat\bbeta_n-\bbeta\|_2^2$. We will see that this change gives an estimator (after a simplification) which obtains by premultiplying the LASSO with the biasing factor, and is a multicollinear resistant estimator.

%Pre-multiplying estimators by the biasing factor $\bF_n(d)$, to improve estimation in risk sense and prediction accuracy, has been used by many. See Akdeniz and Kaciranlar (1995), Akdeniz Duran et al. (2012), and Mansson et al. (2012, 2016) to mention some recent studies. Nevertheless, it is not yet motivated and used with the LASSO.

%Mansson, K., Kibria, B.M. Golam and Shukur, G. (2012) On Liu estimators for the logit regression model, Economic Modeling, 29, 1483-1488.

%Akdeniz, F., Kaciranlar, S. (1995). On the almost unbiased generalized Liu estimator and unbiased estimation of the bias and MSE. Communications in Statistics: Theory and Methods 24:1789–1797.

%Mansson, K., Kibria, B.M. Golam and Shukur, G. (2016) A restricted Liu estimator for binary regression models and its application to an applied demand system, J. Appl. Statist., 43(6), 1119-1127.

%Esra Akdeniz Durana, , Wolfgang Karl Härdle, Maria Osipenko, (2012) Difference based ridge and Liu type estimators in semiparametric regression models, J. Mult. Anal., 105(1), 164-175.

%\subsection{Plan of paper}
%Section 2 contains the proposal of linear unified LASSO as a solution to
In Section 2 we define the linear unified LASSO (LLASSO) and discuss about selecting the biasing parameter a little. In general, we modify the $l_1$-penalty term of LASSO and then propose a closed form solution. In Section 3, we communicate about some asymptotic properties. We show that the LLASSO is $\sqrt{n}$-consistent. Also orthonormal design case is studied. Section 4 is devoted to an extensive numerical study. Two real examples and five simulated examples are considered to compare the performance of LLASSO with the existing candidates including the ridge, LASSO and elastic net, while Section 5 contains conclusions and suggestions for further research. Proofs of all theorems are provided in the Appendix.

\setcounter{equation}{0}
\section{Linear Unified LASSO}
In this section, we propose an estimator called linear unified LASSO (LLASSO) via the penalized least squares approach.
\subsection{Na\"ive look}
Before giving the expression of LLASSO, we first study the effect of replacing $\lambda_2\|\bbeta\|_2^2$ in the E-net by $\lambda_2\|d\hat\bbeta_n-\bbeta\|_2^2$.
As in Zou and Hastie (2005), we assume that the response is centered and the predictors are standardized. For the fixed $\lambda_1$, $\lambda_2$, and $0<d<1$, define the na\"ive loss
\begin{equation*}
L(\bbeta;\lambda_1,\lambda_2,d)=\|\bY-\bX\bbeta\|_2^2+\lambda_2\|d\hat\bbeta_n-\bbeta\|_2^2+\lambda_1\|\bbeta\|_1.
\end{equation*}
The following result gives the solution to the underlying optimization problem in above, similar to Zou and Hastie (2005). 
\begin{proposition}
Suppose $\dot{\bbeta}_n=\arg\min_{\bbeta}L(\bbeta;\lambda_1,\lambda_2,d)$. Then,
\begin{equation*}
\dot{\bbeta}_n=\frac{1}{\sqrt{1+\lambda_2}}\arg\min_{\bb}\calL(\bb;\gamma),
\end{equation*}
where 
\begin{equation*}
\calL(\bb;\gamma)=\|\bY^*-\bX^*\bb\|_2^2+\gamma\|\bb\|_1,\quad \gamma=\frac{\lambda_1}{\sqrt{1+\lambda_2}},
\end{equation*}
with $\bY^*=(\bY^{\top},0^{\top})^{\top}$, $\bX^*=(1+\lambda_2)^{-\frac12}(\bX^{\top},\sqrt{\lambda_2}\bI_p)^{\top}$, and $\bb=\sqrt{1+\lambda_2}(d\hat\bbeta_n-\bbeta)$.
\end{proposition}
The proof is straight and omitted.

The above result shows that the solution to the na\"ive problem, is an augmented LASSO. However, it does not provide a closed form solution with respect to the biasing parameter $d$. Yet, we deliberate more on the use of Proposition 1.
Note that using this result, the $\textit{l}_2$-error bound can be established easily.
Let $\bbeta^o=(\beta_1^o,\ldots,\beta_p^o)^{\top}$ be the ``true parameter value" and $S_o=\{j:\beta_j^o\neq0\}$, as the active set. Then, $s_o=|S_o|$ is termed as the sparsity index of $\bbeta^o$.

By Theorem 11.1 of Hastie et al. (2016), one has the following bound
\begin{equation}\label{error-bound}
\|\dot{\bbeta}_n-\bb^o\|_2^2\leq\frac{6}{\nu}n\sqrt{s_o}\gamma,\quad \forall\;0<d<1,
\end{equation}
where $\bb^o=\sqrt{1+\lambda_2}(d\hat\bbeta-\bbeta^o)$ and $\nu$ is the lower bound of restricted eigenvalues of $\bC$ over an appropriate constraint set. See Eq. (11.13) of Hastie et al. (2016) for more details. The usefulness of the bound \eqref{error-bound} is that one can make the error small by choosing an appropriate $d$, for which $d\hat\bbeta$ is close to $\bbeta^o$. This is more important for the bound of prediction error. Similar to \eqref{error-bound}, one can set up prediction error bound of LLASSO which is dependent to the factor $\gamma^2$. The result of Lederer et al. (2016) can be also applied here.

From Proposition 1, one can also approximate the standard error. Let $\hat{\sigma}^2$ be the estimate of $\sigma^2$. Then, using the result of Osborne et al. (2000), the variance-covariance matrix of $\dot{\bbeta}_n$ has form $(\bC_n^*+\bW^*)^{-1}\bC_n^*(\bC_n^*+\bW^*)^{-1}\hat{\sigma}^2/(1+\lambda_2)$ with
\begin{equation*}
\bC_n^*+\bW^*=
{\bX^*}^{\top}\left(\bI_n+\frac{\be\be^{\top}}{\|\bbeta^*\|_1\|{\bX^*}^{\top}\be\|_\infty}\right)\bX^*
\end{equation*}
where $\bC_n^*={\bX^*}^{\top}\bX^*$, $\be=\bY^*-\bX^*\bb$, and $\|\bbeta\|_\infty=\max_{1\leq j\leq p}|\beta_j|$.

%In the line of Theorem 2 of Zou and Hastie (), we have the following result.
\begin{proposition}\label{theorem2}
Under the assumptions of Proposition 1, given $(\lambda_1,\lambda_2,d)$, we have
\begin{equation*}
\dot{\bbeta}_n=\arg\min_{\bbeta}\left\{\bbeta^\top\left(\frac{{\bX}^\top\bX+\lambda_2\bI_p}{1+\lambda_2}\right)\bbeta
-2\bY^\top\bX\bbeta+\lambda_1\|d\hat\bbeta_n-\bbeta\|_1\right\}.
\end{equation*}
\end{proposition}
Zou and Hastie (2005) interpreted the E-net solution as a rescaled LASSO, which will improve prediction accuracy. Indeed, the term $\left(\frac{{\bX}^\top\bX+\lambda_2\bI_p}{1+\lambda_2}\right)$ is a shrinkage version of $\bX^\top\bX$, which the latter appears in LASSO. Here, the same interpretation is valid, where we replaced $\|\bbeta\|_1$  by $\|d\hat\bbeta_n-\bbeta\|_1$ in LASSO.

Next, we will be considering an approximated closed-form solution to our optimization problem. This will pave the road to define the LLASSO, after some modifications.

\subsection{LLASSO}
Recall the closed-form approximate solution to the optimization problem
\begin{equation*}
\min_{\bbeta}\left\{\frac{1}{n}\|\bY-\bX\bbeta\|^2_2+\lambda\|\bbeta\|_1\right\}
\end{equation*}
has the form $(\bC_n+\lambda\bW^-)^{-1}\bX^\top\bY$, $\bC_n=\bX^\top\bX$, where $\bW^-$ is the generalized inverse of $\bW=\diag(|\hat\beta_j|)$, with $\hat\bbeta_n=(\hat\beta_1,\ldots,\hat\beta_p)$ (see Tibshirani, 1996). After some algebra, the closed-from approximate solution to the problem
\begin{eqnarray*}
\min_{\bbeta}L(\bbeta;\lambda_1,\lambda_2,d)=\min_{\bbeta}\|\bY-\bX\bbeta\|_2^2+\lambda_2\|d\hat\bbeta_n-\bbeta\|_2^2+\lambda_1\|\bbeta\|_1.
\end{eqnarray*}
is given by
\begin{eqnarray}\label{approximate}
&(\bC_n+\lambda_2\bI_p+\lambda_1\bW^-)^{-1}(\bX^\top\bY+d\lambda_2\hat\bbeta_n)\cr
=&(\bC_n+\lambda_2\bI_p+\lambda_1\bW^-)^{-1}(\bC_n+d\lambda_2\bI_p)\hat\bbeta_n
\end{eqnarray}
Let $\lambda_1=\lambda_2=1$. Then, \eqref{approximate} reduces to
\begin{eqnarray}\label{eq23}
&&(\bC_n+\bI_p+\bW^-)^{-1}(\bX^\top\bY+d\hat\bbeta_n)\cr
=&&\bF_n^*(d)\hat\bbeta_n,\quad \bF_n^*(d)=(\bC_n+\bI_p+\bW^-)^{-1}(\bC_n+d\bI_p),
\end{eqnarray}
which is similar to the Liu estimator, except the coefficient $\bF_n(d)$ is replaced by $\bF_n^*(d)$ here. In conclusion, the approximate closed form solution to our problem shows that the effect of penalization due to $l_1$-norm appears in the Liu estimator by the term $\bW^-$. To avoid inefficiency, we suggest to pre-multiply the term $\bF_n(d)$ to the LASSO solution, for the proposal of LLASSO. This proposal can be also interpreted as re-scaling the LASSO estimator to be multicollinear resistant.

Recall that the na\"ive look does not provide a closed form solution with respect to the biasing parameter. In this case, an approximate closed form is of interest. Similar to Tibshirani (1996), one may make use of $\sum b_j^2/|b_j|$, with $\bb=(b_1,\ldots,b_p)^{\top}$, instead of the penalty term $\|\bb\|_1$ to get the LLASSO, say, by a manipulation on \eqref{eq23} as
\begin{equation}\label{LLASSO}
\hat\bbeta_n^{L}(d)=(\bC_n+\bI_{p})^{-1}(\bX^{\top}\bY+d\hat\bbeta_n^{L})=\bF_n(d)\hat\bbeta_n^{L},
\end{equation}
where $0<d<1$ is the biasing parameter and $\bF_n(d)=(\bC_n+\bI_p)^{-1}(\bC_n+d\bI_p)$ is the biasing factor.

\subsection{Choice of biasing parameter}
Apparently, the LLASSO is linear in terms of $d$.
According to \eqref{error-bound}, we seek for such $d$ for which $d\hat\bbeta_n$ is close to $\bbeta^o$. Therefore, one possible choice can be either $\min_d\|d\hat\bbeta_n-\hat\bbeta_n^L\|_1$ or $\min_d\|d\hat\bbeta_n-\hat\bbeta_n^{En}\|_1$. This problem can be solved by an optimization method such as interior point which is of polynomial order.

However, a general formula can be obtained as follows. Solving the loss function $L(\bbeta;\lambda_1,\lambda_2,d)$ with respect to $d$ yields
\begin{equation*}
d=\frac{1}{\lambda_2\hat\bbeta_n^{\top}\hat\bbeta_n}\left\{\lambda_2\hat\bbeta_n^{\top}\bbeta\pm
Q(\bbeta;\lambda_1,\lambda_2)^{\frac12}\right\}
\end{equation*}
where
\begin{equation*}
Q(\bbeta;\lambda_1,\lambda_2)=\lambda_2^2(\hat\bbeta_n^{\top}\bbeta)^2-\lambda_2\hat\bbeta_n^{\top}\hat\bbeta_n L(\bbeta;\lambda_1,\lambda_2)
\end{equation*}
with $L(\bbeta;\lambda_1,\lambda_2)=\|\bY-\bX\bbeta\|_2^2+\lambda_1\|\bbeta\|_1+\lambda_2\|\bbeta\|_2^2$.

If $\bbeta$ is sparse, then $\hat\bbeta_n^{En}\leq\hat\bbeta_n$ and hence
\begin{eqnarray}
\hat Q&=&\max_{\bbeta}Q(\bbeta;\lambda_1,\lambda_2)=\lambda_2^2(\hat\bbeta_n^{\top}\hat\bbeta_n)^2-\lambda_2\hat\bbeta_n^{\top}\hat\bbeta_n L(\hat\bbeta_n^{En};\lambda_1,\lambda_2)\cr
\mbox{and}&&\cr
%\hat d&=&1-\frac{\hat Q^{\frac12}}{\lambda_2\hat\bbeta_n^{\top}\hat\bbeta_n}
\hat d&=& max\left(0,1-\frac{\hat Q^{\frac12}}{\lambda_2\hat\bbeta_n^{\top}\hat\bbeta_n}\right)
\end{eqnarray}

The forthcoming section is devoted to the properties of the LLASSO as defined by \eqref{LLASSO}.
\section{Asymptotic Properties}
In this section, we establish some properties of the LLASSO.

%A basic inequality reads (Lemma 6.1 page 103 of Peter and Sara)

In sequel we will be assuming the following regularity conditions:
\begin{enumerate}
\item[(A1)] $\bC_n=\frac1n\sum_{i=1}^n \bx_i\bx_i^{\top}\to\bC$, $\bC$ is a non-negative definite matrix.
\item[(A2)] $\frac1n\max_{1\leq i\leq n}\bx_i^{\top}\bx_i\to0$.
\item[(A3)] $\bF_n(d)\to\bF(d)$,\quad  $\bF(d)=(\bC+\bI_p)^{-1}(\bC+d\bI_p)$.
\end{enumerate}
For our purpose we assume $\bC$ is nonsingular.
\begin{proposition}\label{theorem3}
Suppose $\bphi=\hat\bbeta_n^L$ is the minimizer of
\begin{equation*}
Z_n(\bphi)=\frac{1}{n}\|\bY-\bX\bphi\|^2_2+\frac{1}{n}\lambda_n\|\bphi\|_1,\quad \bphi=(\phi_1,\ldots,\phi_p)^{\top}
\end{equation*}
Under the set of local alternatives $\calK_{(n)}:\bbeta=\bbeta_{(n)}=\frac{\bdelta}{\sqrt n}$, $\bdelta=(\delta_1,\ldots,\delta_q)\neq\0$, assume (A1)-(A3). If $\lambda/n\to\lambda_o\geq0$, then, we have
\begin{equation*}
\sqrt n(\hat\bbeta_n^{L}(d)-\bbeta)\overset{\mathcal{D}}{\to}\bF(d)\left[\arg\min_{\bu} V(\bu)+\bdelta\right],
\end{equation*}
where $V(\bu)=-2\bu^{\top}\bW+\bu^{\top}\bC\bu+\lambda_o\sum_{j=1}^p \left[u_j\textnormal{sgn}(\beta_j)I(\beta_j\neq0)+|u_j+\delta_j|I(\beta_j=0)\right]$ and $\bW\sim\calN_p(\0,\sigma^2\bC)$.
\end{proposition}

\subsection{Orthonormal design}
Suppose $\bC_n=\bI_p$. Then the LLASSO has form
\begin{eqnarray*}
\hat\beta_{jn}^{L}(d)=c_d\textnormal{sgn}(\hat\beta_j)(|\hat\beta_j|-\lambda/2)^+,\quad j=1,\ldots,p,
\end{eqnarray*}
where $a^+=\max(0,a)$, $\lambda$ is determined by the condition $\sum|\hat\beta_j|=t$, $c_d=(1+d)/2$, and $\hat\beta_j$ is the $j$-th component of OLS estimator.
The estimator $\hat\beta_{jn}^{L}(d)$ is termed normalized LASSO in our terminology. It can be simply verified that the normalized LASSO is the solution of the following optimization problem
\begin{equation}
\min_{\bbeta}\left\{\frac{1}{n}\|\bY-\bX\bbeta\|^2_2+c_d\lambda\|\bbeta\|_1\right\}.
\end{equation}
Under normality assumption, some interesting properties can be achieved. Hence, suppose that the error term in \eqref{eq11} has normal distribution with zero mean and covariance matrix $\sigma^2\bI_n$, where $\sigma^2$ is known. Then, $\hat\beta_j\sim \calN(\beta_j,\sigma^2)$.
\begin{proposition}\label{theorem4} For all $\delta\leq \frac12$ and $\lambda=2\sigma\sqrt{2\log\delta^{-1}}$
\begin{eqnarray*}
\mathbb{E}\left[\hat\beta_{jn}^{L}(d)-\Delta_j\right]^2&\leq&\sigma^2c_d^2(1+2\log\delta^{-1})[\delta+\min(\Delta_j^2,1)]+(\sigma c_d-1)^2\Delta^2_j\cr
&&-2\sigma c_d(\sigma c_d-1)\Delta_j(\lambda/2\sigma)\left[\Phi(\lambda/2\sigma-\Delta_j)-\Phi(\lambda/2\sigma+\Delta_j)\right],
\end{eqnarray*}
where $\Delta_j=\beta_j/\sigma$ and $\Phi(\cdot)$ is the cumulative distribution function of the standard normal distribution.
\end{proposition}
It can be also shown that
\begin{eqnarray}
{\rm Risk}(\hat\bbeta_n^L(d))&=&\mathbb{E}\|\hat\bbeta_n^L(d)-\bbeta\|_2^2\cr
&=&c_d^2\sum_{j=1}^p\left\{1+\lambda_o^2+(\Delta^2_j-1-\lambda_o^2)\left[\Phi(\lambda_o-\Delta_j)-\Phi(-\lambda_o-\Delta_j)\right]\right.\cr
&&\left.-(\lambda_o-\Delta_j)\varphi(\lambda_o+\Delta_j)-(\lambda_o+\Delta_j)\varphi(\lambda_o-\Delta_j)\right\},
\end{eqnarray}
where $\lambda_o=\lambda/2$ and $\varphi(\cdot)$ is the probability density function of the standard normal distribution

%In what follows, we apply the results of Donoho and Johnstone (1994) to obtain some results about the quadratic risk of normalized LASSO.

\section{Numerical Studies}
In this section, we compare the performance of the LLASSO with some other known estimators. 

\subsection{Illustration}
In the following, we study two real life examples. The predictors for each data sets were standardized to have
zero mean and unit standard deviation before fitting the model. We also center the response variable. We then fit linear regression model to predict the variables of
interest using the available regressors. 
We evaluate the performance of the estimators by averaged cross validation (CV) error using a 10-fold CV. In CV, the estimated $\rm MSE_y$ varies across runs. Therefore, we repeat the process 250 times, and calculate the median $\rm MSE_y$ and its standard error. The results are given in Table \ref{Tab:real:results}. Analyzing these results reveal the following conclusions:

\begin{itemize}
\item Regarding the state data: we observe that the LLASSO has the least $\rm MSE_y$ value among all alternatives methods. The second best method is the ridge, having the least standard error.

\item For the prostate data, the ridge estimator has the best performance since there exists the problem of multicollinearity. However, if both multicollinearity and variable section are important, the LLASSO is preferred since it performs better than all others.

\item Surprisingly, the performance of the LLASSO is more efficient compared to the LASSO and E-net.

\end{itemize}

In what follows we only describe the data sets we used.

\subsubsection{State Data}
Faraway (2002) illustrated variable selection methods using the state data set. There are 50 observations (cases) on 8 variables. The variables are: population estimate as of July 1,
1975; per capita income (1974); illiteracy (1970, percent of population); life expectancy in
years (1969-71); murder and non-negligent manslaughter rate per 100,000 population (1976);
percent high-school graduates (1970); mean number of days with minimum temperature 32
degrees (1931-1960) in capital or large city; and land area in square miles. We consider life expectancy as the response (refer to Table 1).

\begin{table}[!htbp]
%\begin{adjustbox}{width=1\textwidth}
\small
\centering
\begin{tabular}{ll}
\hline
\textbf{Variables} & \textbf{Descriptions} \\
\hline \hline

\textbf{Dependent Variable} &\\
lifex & life expectancy in years (1969-71)  \\
{(\bf needed check)} \\
\hline \hline

\textbf{Covariates}   \\
population      & population estimate as of July 1, 1975\\
income          & per capita income (1974) \\
illiteracy      & illiteracy (1970, percent of population) \\
murder          & murder and non-negligent manslaughter rate per \\                 &100,000 population (1976)  \\
hs.grad         & mean number of days with minimum temperature 32\\                 &degrees (1931-1960) in capital or large city  \\
area            & land area in square miles\\
\hline 
\end{tabular}
%\end{adjustbox}
\caption{Descriptions of variables for the state data set
\label{Tab:variables:prostate}}
\end{table}

%\begin{figure}[!htbp]
%\centering
%\includegraphics[width=9cm,height=9cm]{cor_plot.eps}
%\caption{Correlation among predictors for State Data.
% \label{Fig:corplot:PE}}
%\end{figure}

\begin{figure}
    \centering
    \begin{subfigure}[b]{.5\textwidth}
        \centering
        \includegraphics[width=7cm,height=7cm]{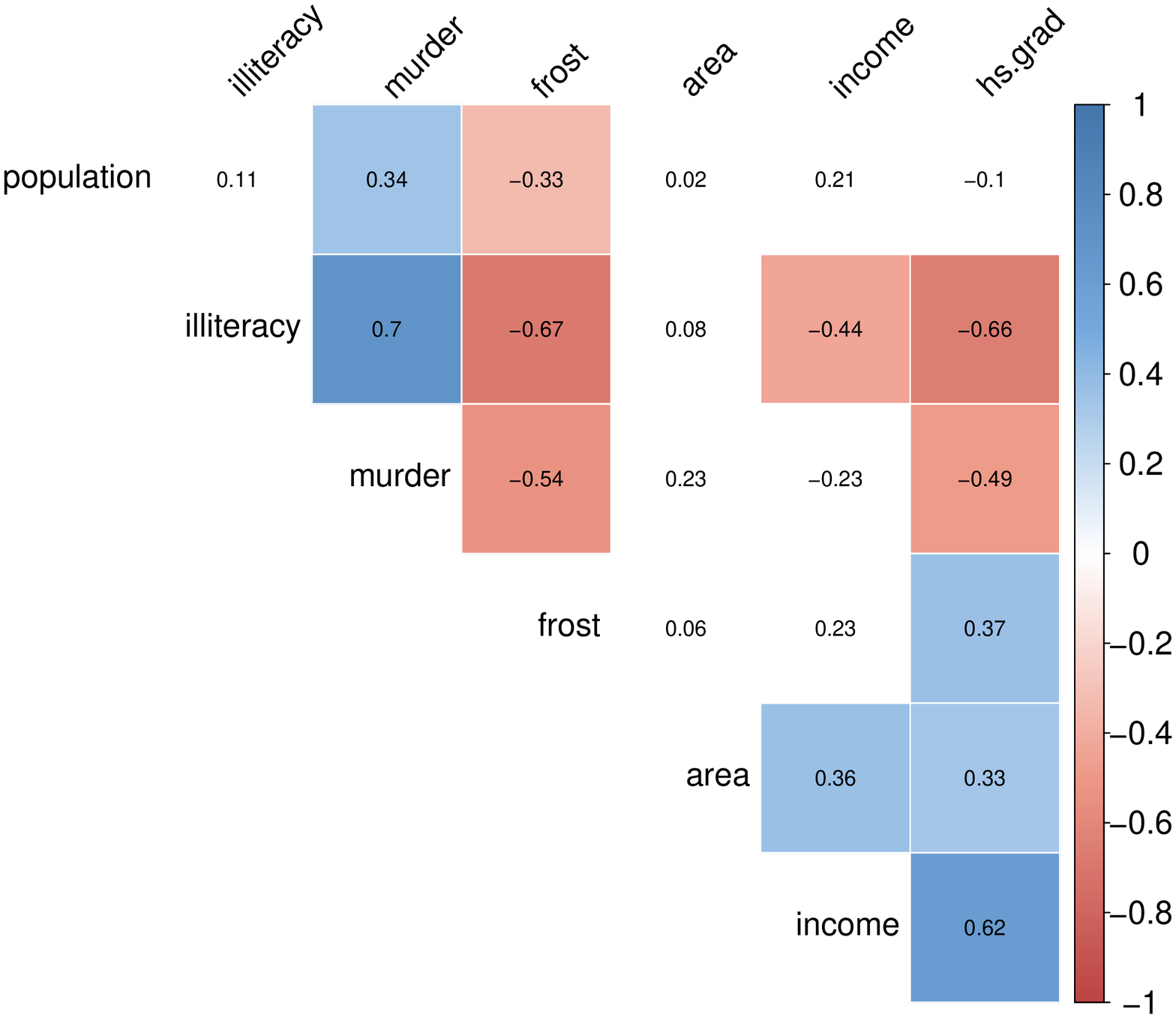}
        \caption{State Data}
        \label{fig:state:cor}
    \end{subfigure}\hfill
    \begin{subfigure}[b]{.5\textwidth}
        \centering
        \includegraphics[width=7cm,height=7cm]{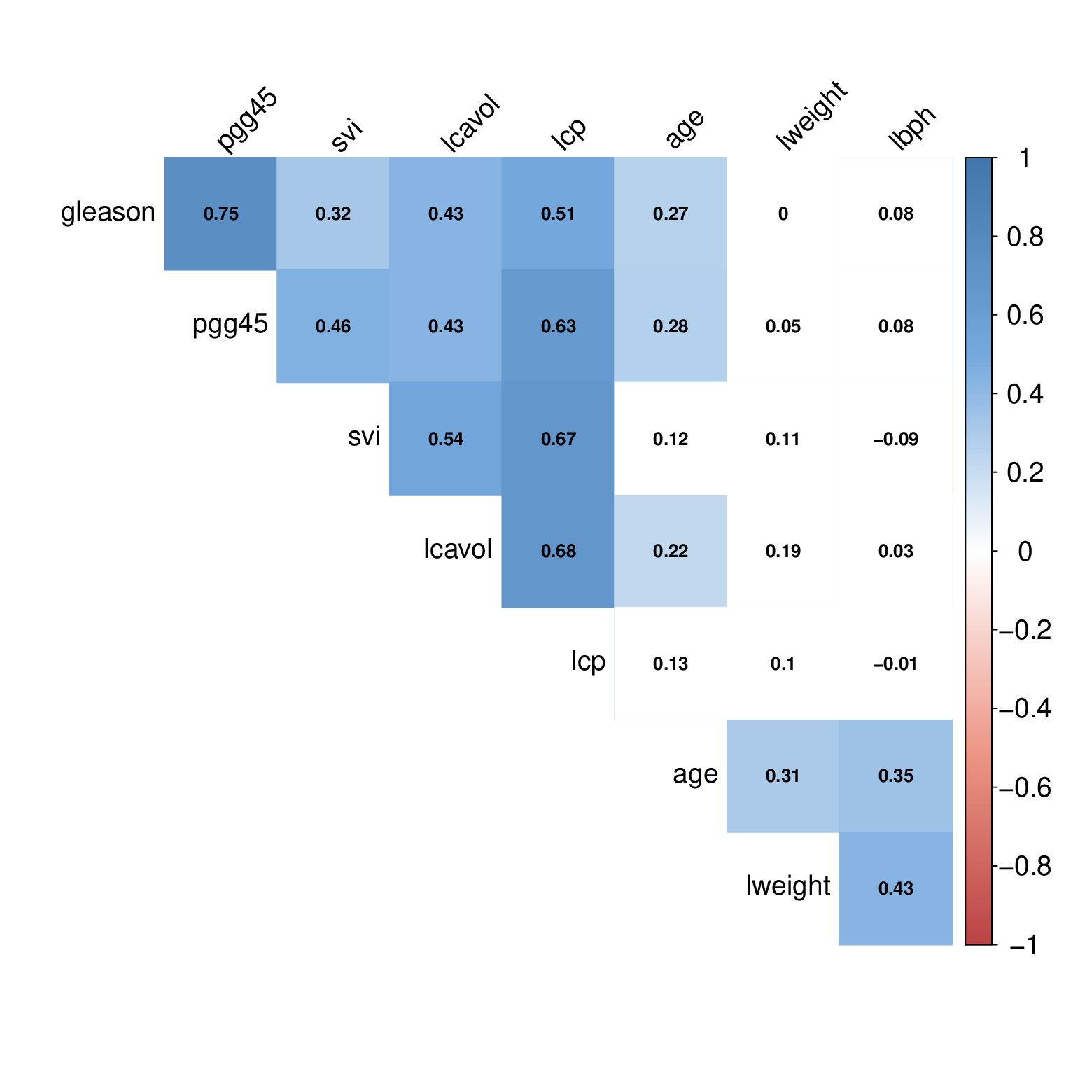}
        \caption{Prostate Data}
        \label{fig:prostate:cor}
    \end{subfigure}
    \caption{Correlations among predictors}
    \label{Deltas}
\end{figure}

%\subsubsection{Diabetes Data}
%In this study, we consider the diabetes data which is used in the study of Efron et al. (2004). The dataset consists of 10 covariates: AGE, SEX, BODY, Body Mass Index (BMI), Blood Pressure (BP), and six blood serum measurements, S1--S6 for each of n = 442 patients. The response variable is a quantitative measure of disease progression after 1 year of follow-up. The objective of this study is to examine the disease progression associated with important prognostic variables. 

\subsubsection{Prostate Data}
Prostate data came from the study of Stamey et al. (1989) about correlation between the level of prostate specific antigen (PSA), and a number of clinical measures in men who were about to receive radical prostatectomy. The data consist of 97 measurements on the following variables: log cancer volume (lcavol), log prostate weight (lweight), age (age), log of benign prostatic hyperplasia amount (lbph), log of capsular penetration (lcp), seminal vesicle invasion (svi), Gleason score (gleason), and percent of Gleason scores 4 or 5 (pgg45). The idea is to predict log of PSA (lpsa) from these measured variables.

A descriptions of  the variables in the prostate dataset is given in Table \ref{Tab:variables:prostate}.

\begin{table}[!htbp]
%\begin{adjustbox}{width=1\textwidth}
\small
\centering
\begin{tabular}{ll}
\hline
\textbf{Variables} & \textbf{Descriptions} \\
\hline \hline

\textbf{Dependent Variable} &\\
lpsa & Log of prostate specific antigen (PSA) \\
\hline \hline

\textbf{Covariates}   \\
lcavol   & Log cancer volume  \\
lweight  & Log prostate weight \\
age      &  Age in years\\
lbph     & Log of benign prostatic hyperplasia amount  \\
svi      & Seminal vesicle invasion  \\
lcp      & Log of capsular penetration  \\
gleason  &  Gleason score\\
pgg45    & Percent of Gleason scores 4 or 5 \\
\hline 
\end{tabular}
%\end{adjustbox}
\caption{Descriptions of variables for the Prostate data set
\label{Tab:variables:prostate}}
\end{table}

\begin{table}[ht]
\centering
\caption{ $\rm MSE_y$ of estimators.}
\begin{adjustbox}{width=1\textwidth}
\begin{tabular}{crrrrrrr}
  \hline
Dataset& OLS&Ridge&Liu&LASSO&LLASSO&E-net \\ 
  \hline
  
State&$0.94867_{0.01207}$ & $0.94083_{0.01096}$ & $0.94131_{0.01162}$ & $0.94349_{0.01199}$ & $\bf 0.93647_{0.01156}$ & $0.94506_{0.01199}$ \\   
  
%Diabetes&$3063.03099_{4.2912}$ & $\bf 3054.76345_{3.99229}$ & $3057.9827_{4.22212}$ & $3062.93912_{4.27284}$ & $3058.93477_{4.21203}$ & $3062.61957_{4.26936}$ \\   

Prostate& $0.63301_{0.00449}$ & $\bf 0.61922_{0.00424}$ & $0.62918_{0.00444}$ & $0.63196_{0.00448}$ & $0.62816_{0.00442}$ & $0.63202_{0.00448}$ \\  

%Prostate&$0.56934_{0.00131}$ & $0.57304_{0.00269}$ & $0.5647_{0.00123}$ & $0.56532_{0.00153}$ & $0.56369_{0.00147}$ & $0.56864_{0.00199}$ \\

%Pollution& $2615.29337_{38.98244}$ & $\bf 1839.91883_{19.87108}$ & $2098.93973_{26.1678}$ & $2521.21674_{36.70513}$ & $2081.47313_{25.4638}$ & $2522.77938_{36.7634}$ \\
  
%Yes&$3515.252_{5.554}$ & $3507.874_{5.445}$ & $3496.000_{5.344}$ & $3506.119_{5.429}$ & $3490.237_{5.247}$ & $3508.013_{5.457}$ \\  
%No&$30659.267_{83.906}$ & $30641.87_{82.78}$ & $29701.611_{48.569}$ & $30645.817_{83.254}$ & $29698.622_{48.406}$ & $30645.735_{83.257}$\\
%Prostate1&$0.56934_{0.00131}$ & $0.57304_{0.00269}$ & $0.5647_{0.00123}$ & $0.56532_{0.00153}$ & $0.56369_{0.00147}$ & $0.56864_{0.00199}$ \\ 
   \hline
\end{tabular}
\end{adjustbox}
\label{Tab:real:results}
\end{table}

\subsection{Simulation}
The purpose of this section is to design a Monte Carlo simulation to show the superiority of LLASSO over the estimators OLS, ridge, Liu, LASSO and E-net.

We used five examples some of which were also considered in Zou and Hastie (2005). All simulations are based on the model 
\begin{equation*}
\bY = \bX\bbeta+\sigma\bold{\epsilon}    
\end{equation*}
where $\bold{\epsilon}\sim N(0,\boldsymbol{I})$. 
In each example, the simulated data contains a training dataset, validation data and an independent test set. We fitted the model only using the training data and the tuning parameters were selected using the validation data. In simulations, we center all variables based on the training data set. Let $\bar{\bx}_{train}=(\bar{x_1}_{, train},\cdots,\bar{x_p}_{, train})$ denote the vector of means of
the training data, $n_{test}$ the number of observations in the test
data set and $\bar{y}_{train}$ the mean over responses in the training data.
Finally, we computed two measures of performance, the test error (mean squared error) $\rm MSE_y = \frac{1}{n_{test}} \textbf{r}_{sim}^{\top}\textbf{r}_{sim}$ where $\textbf{r}_{sim} = \bx_i\bbeta-(\bar{y}_{train}+(\bx_i-\bar{\bx}_{train})^{\top}\bbhat))$ and the mean squared error of the estimation of $\bbeta$ such that $\rm MSE_{\beta} = |\bbhat-\bbeta|^2$ (see Tutz and Ulbricht, 2009). We use the notation $\cdot/\cdot/\cdot$ to describe the number of observations in
the training, validation and test set respectively. Here are the details of five examples:

\begin{itemize}
\item[1-] Each data set consists of $20/20/200$ observations. $\bbeta$ is set to $\bbeta^{\top} = (3, 1.5, 0, 0, 2, 0, 0, 0)$ and $\sigma = 3$. Also, we generate $\bX \sim N(0,\boldsymbol{\Sigma})$, where $\Sigma_{ij}=0.5^{|i-j|}$.

\item[2-] Each data set consists of $100/100/300$ observations and $40$ predictors, where $\beta_j=0$ when $j=1,\dots,10, 21,\dots,30$ and $\beta_j=3$ when $j=11,\dots,20, 31,\dots,40$. Also, we set $\sigma=3$, $\Sigma_{ij}=0.5^{|i-j|}$, as in example 1. 
%for $i\neq j$ and $\Sigma_{ij}= 1$ for $i = j$.

\item[3-]  Each data set consists of 50/50/200 observations and 30 predictors. This setting was also considered in El Anbari and Mkhdari (2014) with a slight change. 
We chose
\begin{equation*}
\bbeta =\left( \underbrace { 3,\dots,3}_{5}, \underbrace {4,\dots,4}_{5},\underbrace {0,\dots,0}_{20} \right) 
\end{equation*}
and $\sigma=3$. The predictors $\bX$ were generated as follows
\begin{align*}
\bx_i =& Z_1+\varepsilon^x_i, Z_1\sim \mathcal{N}(0,1), i=1,\dots,5,\\
\bx_i =& Z_2+\varepsilon^x_i, Z_2\sim \mathcal{N}(0,1), i=6,\dots,10, \\
\bx_i \sim& \mathcal{N}(0,1), i=11,\dots,30. 
\end{align*}

\item[4-] Each data set consists of $20/20/200$ observations. $\bbeta$ is specified by $\bbeta^{\top} = (3, 1.5, 0, 0, 0, 0, -1, -1)$ so that there are two positively and two negatively correlated predictors which are truely relevant and $\sigma = 3$. We also consider $\bX \sim N(0,\boldsymbol{\Sigma})$, where $\Sigma_{ij}=0.5^{|i-j|}$.

\item[5-] Each data set consists of 50/50/200 observations and 30 predictors.
We chose
\begin{equation*}
\bbeta =\left( \underbrace { 2,\dots,2}_{8},\underbrace {0,\dots,0}_{22} \right). 
\end{equation*}
Also, we consider $\sigma = 6$ and $\bX \sim N(0,\boldsymbol{\Sigma})$, where $\Sigma_{ij}=0.9^{|i-j|}$.

\end{itemize}

\begin{table}[!htbp]
\centering
%\small
  \caption{Median mean-squared errors for the simulated examples and
five methods based on 250 replications$^*$}
  \label{table1}
\begin{adjustbox}{width=1\textwidth}
\begin{tabular}{lcccccccccccccc}
\\[-1.8ex]\hline
\hline \\[-1.8ex]
&\multicolumn{2}{c}{Example 1} &\multicolumn{2}{c}{Example 2}&\multicolumn{2}{c}{Example 3}&\multicolumn{2}{c}{Example 4}&\multicolumn{2}{c}{Example 5} \\
\cmidrule(lr){2-3} \cmidrule(lr){4-5} \cmidrule(lr){6-7} \cmidrule(lr){8-9} \cmidrule(lr){10-11}
& Median& Median& Median& Median& Median& Median& Median& Median& Median& Median\\ 
&$\rm MSE_y$ &$\rm MSE_{\beta}$ &$\rm MSE_y$ &$\rm MSE_{\beta}$&$\rm MSE_y$ &$\rm MSE_{\beta}$&$\rm MSE_y$ &$\rm MSE_{\beta}$&$\rm MSE_y$ &$\rm MSE_{\beta}$ \\
\\[-1.8ex]\hline
\hline \\[-1.8ex]
OLS & $5.723_{0.319}$ & $8.941_{0.568}$ & $6.980_{0.134}$ & $10.521_{0.216}$ & $69.039_{1.210}$ & $36.079_{0.741}$ & $5.625_{0.316}$ & $8.576_{0.566}$ & $49.400_{1.429}$ & $441.903_{14.301}$ \\ 
 Ridge & $3.494_{0.177}$ & $4.439_{0.190}$ & $5.702_{0.108}$ & $7.289_{0.130}$ & $49.000_{0.795}$ & $24.415_{0.362}$ & $3.457_{0.169}$ & $4.038_{0.178}$ & $7.799_{0.332}$ & $\bf 13.768_{0.976}$ \\ 
Liu & $4.713_{0.240}$ & $6.591_{0.347}$ & $6.707_{0.127}$ & $9.940_{0.200}$ & $62.336_{1.076}$ & $31.896_{0.595}$ & $4.510_{0.217}$ & $6.116_{0.324}$ & $23.271_{0.630}$ & $155.465_{5.395}$ \\ 
  LASSO & $3.225_{0.182}$ & $4.025_{0.239}$ & $\bf 4.668_{0.102}$ & $5.876_{0.135}$ & $46.347_{0.812}$ & $20.873_{0.373}$ & $3.187_{0.163}$ & $3.815_{0.189}$ & $8.083_{0.335}$ & $26.158_{1.250}$ \\ 
LLASSO & $\bf 3.017_{0.172}$ & $\bf 3.652_{0.184}$ & $4.680_{0.101}$ & $5.687_{0.127}$ & $\bf 43.321_{0.778}$ & $\bf 19.639_{0.342}$ & $\bf 3.059_{0.158}$ & $\bf 3.409_{0.147}$ & $\bf 7.221_{0.326}$ & $15.965_{0.778}$ \\ 
  E-net & $3.073_{0.164}$ & $3.907_{0.186}$ & $4.764_{0.102}$ & $\bf 5.665_{0.124}$ & $44.576_{0.738}$ & $20.041_{0.324}$ & $3.065_{0.159}$ & $3.681_{0.178}$ & $7.443_{0.329}$ & $ 15.897_{1.014}$ \\ 
 % Liu-Enet & $3.039_{0.161}$ & $3.586_{0.155}$ & $4.764_{0.103}$ & $5.563_{0.119}$ & $41.688_{0.708}$ & $18.641_{0.303}$ & $3.002_{0.156}$ & $3.283_{0.14}$ & $6.968_{0.324}$ & $10.905_{0.704}$ \\ 
\hline \\[-1.8ex]
\end{tabular}
\end{adjustbox}
\footnotesize{$^{\ast}$ The numbers in smaller font are the corresponding standard errors of the MSE.}\\
\end{table}

\begin{figure}[!htbp]
\centering
\includegraphics[width=14cm,height=20cm]{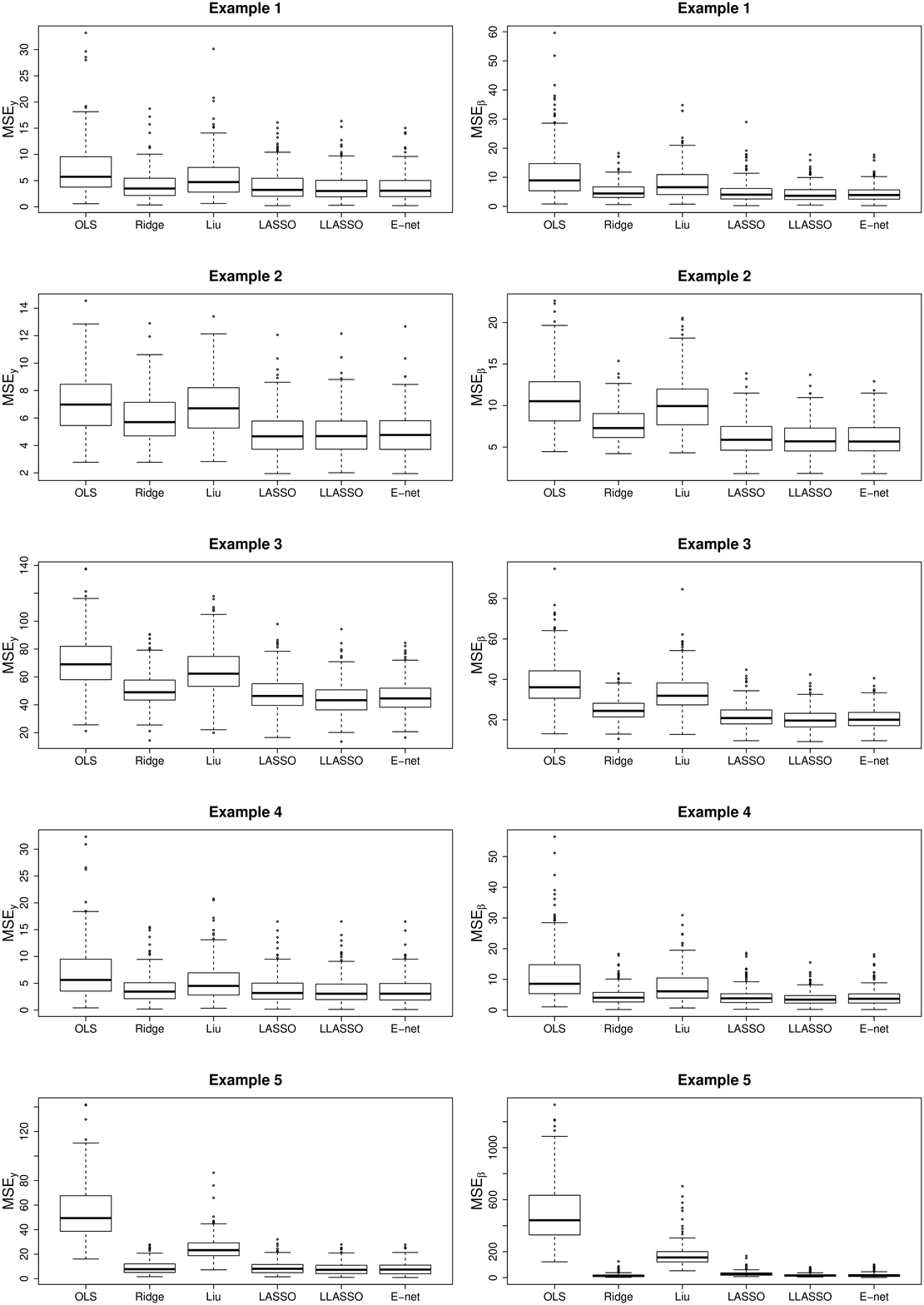}
\caption{Boxplots of test mean squared errors $\rm MSE_y$ (left column) and $\rm MSE_{\beta}$ (right column) for examples 1-5.
 \label{Fig:boxplot:PE}}
\end{figure}

We investigate these scenarios by simulating $250$ data sets. 
The results of the simulation are given in Table~\ref{table1}. We also summarize the results in Figure~\ref{Fig:boxplot:PE} in which we present the box-plots of test mean squared errors $\rm MSE_y$ (left column) and $\rm MSE_{\beta}$ (right column) for examples 1-5. Now, we share the results obtained from the simulation study as follows:

In example 1 with positively correlated variables, although the performances of the estimators are close to each other, LLASSO has the best performance in the sense of both measures. 

In example 2, the LASSO is better compared to all others in the sense of first measure and E-net is the best according to the second criteria. 

In both examples 3 and 4, LLASSO performs better than the others in the sense of both criteria. 

In example 5, we consider the design matrix having the problem of multicollinearity such that the correlations between the predictors are chosen to be $0.9$, and the beta coefficients are sparse. Not surprisingly, the ridge estimator performs better than LASSO while E-net beats the ridge. On the other hand, the performance of LLASSO outshines all others for first measure while ridge is the best for second measure. The LLASSO is competitive with E-net.

%\clearpage
\section{Conclusion}
In this article, we have proposed a new estimator for simultaneous estimation and variable selection. Indeed, we pre-multiplied the LASSO with a matrix factor to become multicollinear resistance, after modifying the $l_1$-norm of the LASSO. The proposed linear unified LASSO or LLASSO for short, has simple form and can be considered as a re-scaled LASSO estimator. The LLASSO inherits all good properties of the LASSO and it is $\sqrt{n}$-consistent. Apart from its good properties, e.g. producing sparse model with good prediction accuracy, there is no need to propose a specific algorithm for its computation. Similar to adaptive LASSO, the LLASSO can be solved by the same efficient algorithm for solving the LASSO.
According to the numerical findings, we suggest to use LLASSO estimation method in practical examples.

For further research, it can be suggested to pre-multiply the term $\bF_n(d)$ to the relaxed LASSO (Meinshausen, 2007) for a faster convergence rate. Any sparse solution of high-dimensional problems can be also substituted with LASSO in our methodology. To construct an estimator with oracle properties, we suggest to use the adaptive LASSO instead of LASSO in the LLASSO. One can also designate the generalized LLASSO. It is defined by
\begin{equation}\label{GLLASSO}
\hat\bbeta_n^{LL}=(\bC_n+\bI_{p})^{-1}(\bX^{\top}\bY+\bD\hat\bbeta_n^{L})=\bF_D\hat\bbeta_n^{L},
\end{equation}
where $\bD=\diag(d_1,\ldots,d_p)$ is the biasing matrix and $\bF_D=(\bC_n+\bI_p)^{-1}(\bC_n+\bD\bI_p)$ is the biasing factor. The generalized LLASSO allows different biasing parameters.

\section*{Appendix: Proofs}
		
\noindent\textbf{Proof of Proposition \ref{theorem2}}

Under the assumptions of Proposition 1, we get
\begin{eqnarray*}
\dot\bbeta_n&=&\arg\min_{\bb}\left\{\left\|\bY^*-\bX^*\frac{\bb}{\sqrt{1+\lambda_2}}\right\|^2_2
+\frac{\lambda_1}{\sqrt{1+\lambda_2}}\left\|\frac{\bb}{\sqrt{1+\lambda_2}}\right\|_1\right\}\cr
&=&\arg\min_{\bbeta}\left\{\left\|\bY^*-\bX^*\frac{(d\hat\bbeta-\bbeta)}{\sqrt{1+\lambda_2}}\right\|^2_2
+\frac{\lambda_1}{\sqrt{1+\lambda_2}}\left\|\frac{(d\hat\bbeta-\bbeta)}{\sqrt{1+\lambda_2}}\right\|_1\right\}\cr
&=&\arg\min_{\bbeta}\left\{(d\hat\bbeta_n-\bbeta)^\top\frac{{\bX^*}^\top\bX^*}{1+\lambda_2}(d\hat\bbeta_n-\bbeta)-2\frac{{\bY^*}^\top\bX^*\bbeta}{\sqrt{1+\lambda_2}}
+{\bY^*}^\top\bY^*\right.\cr
&&\left.\frac{\lambda_1\|d\hat\bbeta_n-\bbeta\|_1}{1+\lambda_2}\right\}\cr
&=&\arg\min_{\bbeta}\left\{\bbeta^\top\left(\frac{{\bX}^\top\bX+\lambda_2\bI_p}{1+\lambda_2}\right)\bbeta
-2\bY^\top\bX\bbeta+\lambda_1\|d\hat\bbeta_n-\bbeta\|_1\right\}.
\end{eqnarray*}
The proof is complete.
\hfill$\blacksquare$

\noindent\textbf{Proof of Proposition \ref{theorem3}}

Note that $\sqrt n(\hat\bbeta_n^{L}(d)-\bbeta)=\sqrt n\bF_n(d)(\hat\bbeta_n^L-\bbeta)+\sqrt n(\bF_n(d)-\bI_p)\bbeta$. Under $\calK_{(n)}$ and (A3), $\sqrt n(\bF_n(d)-\bI_p)\bbeta\to\bF(d)\bdelta$. Also, using Theorem 2 of Knight and Fu (2000), $\sqrt n(\hat\bbeta_n^{L}-\bbeta)\overset{\mathcal{D}}{\to}\arg\min_{\bu} V(\bu)$. Then, the result follows from Slutsky's theorem.
\hfill$\blacksquare$

\noindent\textbf{Proof of Proposition \ref{theorem4}}

Let $Z_j=\hat\beta_j/\sigma$. Then $Z_j\sim\calN(\Delta_j,1)$, $\Delta_j=\beta_j/\sigma$, and we have
\begin{eqnarray*}
\hat\beta_{jn}^{L}(d)=\sigma c_d\textnormal{sgn}(Z_j)(|Z_j|-\lambda/2\sigma)^+
\end{eqnarray*}
Therefore
\begin{eqnarray*}
\mathbb{E}\left[\hat\beta_{jn}^{L}(d)-\Delta_j\right]^2&=&\sigma^2c_d^2\mathbb{E}\left[\textnormal{sgn}(Z_j)(|Z_j|-\lambda/2\sigma)^+-\Delta_j\right]^2+(\sigma c_d-1)^2\Delta^2_j\cr
&&+2\sigma c_d(\sigma c_d-1)\Delta_j \mathbb{E}\left[\textnormal{sgn}(Z_j)(|Z_j|-\lambda/2\sigma)^+-\Delta_j\right]
\end{eqnarray*}
After some algebra
\begin{eqnarray}\label{eq1}
\mathbb{E}\left[\textnormal{sgn}(Z_j)(|Z_j|-\lambda/2\sigma)^+\right]&=&
\mathbb{E}\left[\textnormal{sgn}(Z_j)(|Z_j|-\lambda/2\sigma)I(|Z_j|>\lambda/2\sigma)\right]\cr
&=&\mathbb{E}\left[Z_jI(|Z_j|>\lambda/2\sigma)\right]-\mathbb{E}\left[(\lambda/2\sigma)\textnormal{sgn}(Z_j)I(|Z_j|>\lambda/2\sigma)\right]\cr
&=&\Delta_j-(\lambda/2\sigma)\left[\Phi(\lambda/2\sigma-\Delta_j)-\Phi(\lambda/2\sigma+\Delta_j)\right]
\end{eqnarray}
On the other hand, using Theorem 1 of Donoho and Johnstone (1994)
\begin{eqnarray}\label{eq2}
\mathbb{E}\left[\textnormal{sgn}(Z_j)(|Z_j|-\lambda/2\sigma)^+-\Delta_j\right]^2\leq(1+2\log\delta^{-1})[\delta+\min(\Delta_j^2,1)]
\end{eqnarray}
Using \eqref{eq1} together with \eqref{eq2} yield
\begin{eqnarray*}
\mathbb{E}\left[\hat\beta_{jn}^{L}(d)-\Delta_j\right]^2&\leq&\sigma^2c_d^2(1+2\log\delta^{-1})[\delta+\min(\Delta_j^2,1)]+(\sigma c_d-1)^2\Delta^2_j\cr
&&-2\sigma c_d(\sigma c_d-1)\Delta_j(\lambda/2\sigma)\left[\Phi(\lambda/2\sigma-\Delta_j)-\Phi(\lambda/2\sigma+\Delta_j)\right]
\end{eqnarray*}
which completes the proof.
\hfill$\blacksquare$

\section*{References}
\baselineskip=12pt
\def\ref{\noindent\hangindent 25pt}

\ref A. E. Hoerl, R.W. Kennard, Ridge regression: Biased estimation for nonorthogonal problems. Technometrics, 12(1) (1970) 55-67.

\ref  A.N. Tikhonov, "Solution of incorrectly formulated problems and the regularization method" Soviet Math. Dokl., 4 (1963) 1035–1038 MR0211218 Zbl 0141.11001

\ref  H. Zou, T. Hastie, Regularization and variable selection via the elastic net. J. Royal. Statist. {\rm B}, 67(1) (2005) 301-320.

\ref J. Lederer, L. Yu, I. Gaynanova, Oracle inequalities for high-dimensional prediction, arXiv:1608.00624v1 [math.ST] 1 Aug 2016.

\ref K. Knight, W. Fu, Asymptotics for LASSO-type estimators, Ann. Statist. 28 (2000) 1356-1378.

\ref L.S. Mayer, T. A.Willke, On biased estimation in linear models, Technometrics 15 (1973) 497-508.

\ref N. Meinshausen, Relaxed Lasso, Comp. Statist. Data Anal., 52 (2007) 374-393.

\ref M. El Anbari, A. Mkhadri, Penalized regression combining the L1 norm and a correlation based penalty, The Indian Journal of Statistics 76-B, Part 1 (2008) 82-102.

\ref M.R. Osborne, B. Presnell, B.A. Turlach,  On the LASSO and its Dual, J. Comp. Graph. Statist., 9(2) (2000) 319-337.

\ref R. Tibshirani, Regression shrinkage and selection via the LASSO, J. Royal. Statist. {\rm B}, 58(1) (1996) 267-88.

\ref T. Hastie, R. Tibshirani, M. Wainwright (2016) Statistical Learning with Sparsity The Lasso and Generalizations, Chapman \& Hall/CRC Press.

%\bibitem[McDonald and Schwing, 1973]{McDonald-Schwing1973}
%\textsc{McDonald, G.C..\ \textup{and} Schwing, R.C.} (1973). 
%Instabilities of regression estimates relating air pollution to %mortality. 
%\textit{Technometrics},  
%\textbf{15}(3), 463--481.

\end{document}